\documentclass{article}
\usepackage{amsmath,graphicx}
\usepackage[preprint]{spconf}

\copyrightnotice{\copyright\ IEEE 2024}
\toappear{To appear in {\it Proc.\ ISBI 2024,
                   May 27-30, 2024, Athens, Greece}}

\usepackage{enumitem}
\setlist{nosep, leftmargin=14pt}

\title{Interpretable Models for Detecting and Monitoring Elevated Intracranial Pressure}
\name{%
\begin{tabular}{@{}c@{}}
Darryl Hannan\textsuperscript{\rm 1} \qquad
    Steven C. Nesbit\textsuperscript{\rm 1} \qquad
    Ximing Wen\textsuperscript{\rm 1} \qquad
    Glen Smith\textsuperscript{\rm 1} \qquad
    Qiao Zhang\textsuperscript{\rm 1} \\
    Alberto Goffi\textsuperscript{\rm 2} \qquad
    Vincent Chan\textsuperscript{\rm 2} \qquad
    Michael J. Morris\textsuperscript{\rm 3} \qquad
    John C. Hunninghake\textsuperscript{\rm 3}  \\
    Nicholas E. Villalobos\textsuperscript{\rm 3} \qquad
    Edward Kim\textsuperscript{\rm 1} \qquad
    Rosina O. Weber\textsuperscript{\rm 1} \qquad
    Christopher J. MacLellan\textsuperscript{\rm 4}
\end{tabular}}
\address{\textsuperscript{\rm 1}Drexel University, Philadelphia, PA, USA\\
    \textsuperscript{\rm 2}University of Toronto, Toronto, ON, CA\\
    \textsuperscript{\rm 3}Brooke Army Medical Center, Fort Sam Houston, TX, USA\\
    \textsuperscript{\rm 4}Georgia Institute of Technology, Atlanta, GA, USA}
\begin{document}
%\ninept
%
\maketitle
\begin{abstract}
Detecting elevated intracranial pressure (ICP) is crucial in diagnosing and managing various neurological conditions. These fluctuations in pressure are transmitted to the optic nerve sheath (ONS), resulting in changes to its diameter, which can then be detected using ultrasound imaging devices. However, interpreting sonographic images of the ONS can be challenging. In this work, we propose two systems that actively monitor the ONS diameter throughout an ultrasound video and make a final prediction as to whether ICP is elevated. To construct our systems, we leverage subject matter expert (SME) guidance, structuring our processing pipeline according to their collection procedure, while also prioritizing interpretability and computational efficiency. We conduct a number of experiments, demonstrating that our proposed systems are able to outperform various baselines. One of our SMEs then manually validates our top system's performance, lending further credibility to our approach while demonstrating its potential utility in a clinical setting.
\end{abstract}
\begin{keywords}
Machine Learning, Computer Vision, Biomedical Imaging
\end{keywords}

{\footnotesize© 2024 IEEE.  Personal use of this material is permitted.  Permission from IEEE must be obtained for all other uses, in any current or future media, including reprinting/republishing this material for advertising or promotional purposes, creating new collective works, for resale or redistribution to servers or lists, or reuse of any copyrighted component of this work in other works.}

\section{Introduction}
\label{sec:intro}

\begin{figure}[t]
    \centering
    \includegraphics[width=0.8\columnwidth]{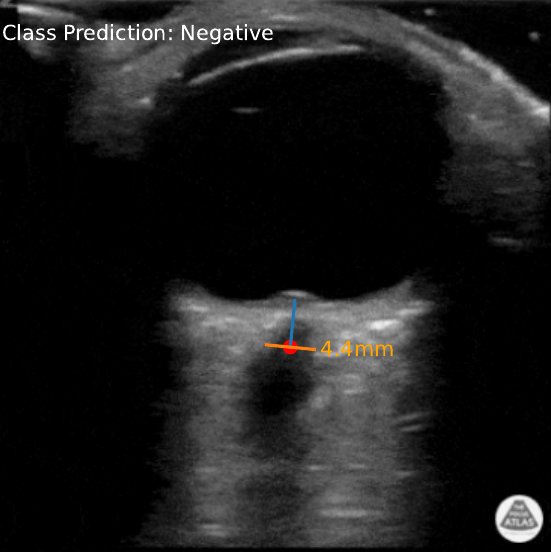}
    \caption{Output from our R2U-Net-based ONSD measurement system. Each stage of our proposed pipeline surfaces clinically relevant information (colored features rendered on the image) that the model is using to make its final prediction.}
    \label{fig:unet_ex}
\end{figure}

Timely and accurate detection of elevated intracranial pressure (ICP) is crucial in diagnosing and managing various neurological conditions, including traumatic brain injury (TBI), hydrocephalus, and more. In the past, invasive ICP monitoring was the sole effective method for detecting elevated ICP. However, this approach necessitates neurosurgical intervention and is associated with significant complications. Moreover, it is not feasible in resource-limited settings, such as remote areas, pre-hospital settings, and facilities without neurosurgical capabilities.

Changes in intracranial pressure are transmitted to the optic nerve sheath (ONS), resulting in corresponding fluctuations in its diameter. By utilizing imaging technologies, it is possible to measure the optic nerve sheath diameter (ONSD), offering a non-invasive approach to detect elevated ICP. Ultrasound holds advantages over other imaging techniques as it can be easily performed at the point-of-care, provides real-time results, is radiation-free, and is cost-effective.
However, acquisition of sonographic images of the ONS can be challenging because the probe needs to be positioned correctly for the nerve and its sheath to be acceptable for measurement. A partially visible nerve and sheath can result in an inaccurate measurement, potentially misdiagnosing the patient.

Due to the challenging nature of this task and the large amount of expertise it requires, it is desirable to have artificial intelligence (AI) systems that can facilitate the procedure. In order for such systems to be effective in these settings, they must meet a few crucial target criteria. First, they need to be designed with the goal of supporting the medical professional rather than replacing them. This involves supporting the entire procedure rather than just making a final diagnosis, and it emphasizes aspects of the system outside of just accuracy, such as interpretability and ease of use. Second, the systems need to compliment the increasing portability of ultrasound devices.
Lastly, labeling medical images is costly due to the high level of expertise required. AI systems need to be capable of achieving strong performance with limited labeled training examples.

In this work, we present two models for elevated ICP detection within the context of an ultrasound video classification task. The first model is a 2D convolutional LCA sparse coding model inspired by the work of Hannan et al. \cite{hannan2022mobileptx}.
The second model uses a R2U-Net \cite{alom2018recurrent} pipeline to directly predict the width of the nerve by generating a mask at the point of measurement (Figure \ref{fig:unet_ex}).\footnote{Image was acquired from the POCUS Atlas (https://www.thepocusatlas.com/).} Both of these models utilize architectures, or are built upon architectures, that have been demonstrated to be effectively run on mobile platforms. We conduct a full set of experiments for both of our proposed models, comparing them to CLIP ViT-B/16 \cite{radford2021learning} and ConvNeXT \cite{liu2022convnet}, two pre-trained models that are exposed to millions of examples. We demonstrate that our proposed models outperform each of these architectures, with our R2U-Net approach exceeding the other models by a large margin. We also conduct a qualitative analysis of this model where a SME assessed its generated predictions, further validating its performance.

While existing approaches have explored measuring individual frames containing the ONS \cite{stevens2021automated,meiburger2021automatic} or determining whether a frame is appropriate for measurement \cite{singh2022good}, an end-to-end system that considers the entire procedure has not been explored previously. Additionally, prior work has not sought to explicitly leverage techniques that are well suited to mobile computing devices, potentially limiting their applicability to a variety of real-world scenarios, nor has it explicitly focused on developing models for limited data settings.

\section{Methodology}
To develop our models, we collaborated with our SMEs, understanding the steps that they follow throughout the procedure. Then we distilled this into our systems by explicitly constructing our processing pipeline to mimic these steps. To briefly summarize, the examiner collects a video, trying to get the nerve into focus by performing small adjustment movements. Once a video containing a properly imaged optic nerve and its sheath has been collected, its careful review will determine which frame(s) demonstrate the OSND and display it accurately for measurement. At this point, a plane located 3 mm posterior to the ocular globe will be identified, specifically targeting the bulbous section of the optic nerve. The objective is to measure the distance of decreased echogenicity between the clearly visible hyperechoic demarcations of the nerve sheath. Various published studies have explored the identification of elevated ICP, often using a cutoff value of 5 mm. If it is found that the distance exceeds this threshold, the examiner may diagnose the patient with elevated ICP.

\subsection{Convolutional Sparse Coding System}
Our first model is inspired by Hannan et al. \cite{hannan2022mobileptx}. The first step in this system is to identify the nerve and determine which frame(s) are suitable for measurement. To do so, we collected an additional 220 labels in the form of bounding boxes around the nerve, then trained a YOLOv5 model \footnote{https://github.com/ultralytics/yolov5} to identify the nerve. YOLOv5 facilitates diagnosis in multiple ways. First, by extracting the nerve and its sheath, it reduces the input that we are passing to subsequent stages of our model. Second, it also allows us to visualize the predicted bounding box for the examiner, increasing the interpretability of the model. Third, it serves as a means of filtering frames that are unfit for measurement because the YOLO model was trained on high quality labeled frames and will be less likely to detect nerves that are only partially visible.

The second step is to sparse code the extracted nerve region using a 2D convolutional variant of the LCA sparse coding algorithm \cite{olshausen1997sparse}. This is a biologically-inspired unsupervised algorithm that is capable of learning a robust set of features given only limited examples. It has previously been demonstrated that such as model can be run in a timely manner on a mobile device \cite{hannan2022mobileptx} and that LCA sparse coding can be executed in an even more efficient manner on neuromorphic hardware\cite{parpart2023implementing}. The sparse coding model is trained on all of the frames available in our ONSD dataset, then at test time just the extracted nerve regions are passed in.
 
Lastly, after applying sparse coding, we train a small convolutional classifier on a binary prediction task, corresponding to whether or not the nerve sheath is over the 5mm threshold. To make a video level prediction, we stride over frames in the video using a fixed interval, get predictions for each frame, average them, and round the result.

\subsection{R2U-Net System}
As an alternative to the sparse coding approach, we further decompose the task after object detection with the goal of increasing the accuracy and interpretability of the model. We collected bounding boxes for the ocular globe and retrained the YOLOv5 model to identify both the ocular globe and ONS. The pipeline then undergoes a number of steps without any machine learning, closely following the procedure that a human examiner would perform to take the measurement. A line is drawn between the center of the ONS and ocular globe, determining the angle of measurement. We measure a point, on this line, that is 3mm from the retinal plane. We crop a 16x128 region that is centered on this point and adjusted to the angle that we calculated previously. We then use a R2U-Net model \cite{alom2018recurrent} to predict a mask that covers the nerve in this region. We use this mask to directly measure the distance between the predicted boundaries of the ONS at the measurement point. The R2U-Net model is trained on 224 masks that an author labeled. These frames were then further curated by discarding any masks that led to a measurement that conflicted with the final video-level label. While the model is only trained on this subset of frames, we use it for video prediction for all available videos, where the procedure is run at fixed intervals over the course of the video, the widths are averaged, and the final video-level prediction is made according to whether the average width exceeds a fixed threshold.

\section{Experiments and Results}
We present results on an ONSD video classification task, comparing to two baselines: CLIP ViT-B/16 \cite{radford2021learning} and ConvNeXT \cite{liu2022convnet}, high performing image classification models using transformers and convolutional networks, respectively. We report both video and frame-level accuracy, where the former is the accuracy of the model in predicting the video-level ground truth classification regarding whether the ONS diameter exceeds 5mm and the latter is the accuracy of the prediction for each individual frame compared to the same ground truth video-level label.

\begin{table}[t]
    \centering
    \begin{tabular}{|c|c|c|}
        \hline
        Model & Video Accuracy & Frame Accuracy \\\hline
        ConvNeXT & 62.00\% $\pm$ 2.65 & 66.34\% $\pm$ 2.52 \\
        CLIP ViT-B/16 & 61.00\% $\pm$ 2.00 & 63.00\% $\pm$ 1.00 \\
        Sparse Coding & 69.00\% $\pm$ 5.20 & 75.00\% $\pm$ 3.46 \\
        R2U-Net & 82.67\% $\pm$ 3.06 & 81.67\% $\pm$ 3.21 \\\hline
    \end{tabular}
    \caption{Accuracy (over 3 runs) for the ONSD dataset.}
    \label{tab:onsd_results}
\end{table}

% \subsection{Experimental Setup}
% 61 videos, 22 positive, 39 negative, 15.30 seconds long
We train the models on an ONSD dataset that was collected by an independent group of SMEs. This dataset contains just 61 videos (22 positive and 39 negative) that are, on average, 15.3 seconds long. Two physicians with expertise in ONSD measurement with ultrasound analyzed each video, determined which frame in the video should be measured, and measured the width of the ONS. If the width of the ONS was larger than 5mm, then the video was classified as `positive', indicating the potential presence of elevated ICP. If the width was less than 5mm, then the video is classified as `negative', indicating that it is unlikely the ICP is elevated.

With the additional labeling that we performed for training our R2U-Net system, we were able to identify high quality frames in the dataset. We trained our sparse coding model on just this subset of data as well and found that it performed better than when the model was trained on all of the available frames. Therefore, this subset is used for all model training but evaluation is done on frames sampled at a fixed interval from each video in the test set. All models are evaluated using 10-fold evaluation and participant grouping, where there is no patient overlap between the training and test sets.

\subsection{Results}
\begin{figure}[t]
    \centering
    \includegraphics[width=0.85\columnwidth]{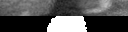}
    \caption{Grouth truth nerve slice (top) and predicted nerve mask (bottom) for POCUS Atlas image.}
    \label{fig:stacked_nerve}
\end{figure}

Table \ref{tab:onsd_results} contains the results of our experiments. Our proposed R2U-Net architecture obtains 82.67\% video accuracy while the sparse coding model obtains 69\%. This gap in performance illustrates the benefits of further task decomposition, even at the cost of additional labeling. The ConvNeXT model and the ViT-B/16 model only obtain 62\% and 61\%, respectively. This illustrates that despite the models being pre-trained on large datasets, fine-tuning them on a small medical dataset is challenging; they are not well suited for the task out of the box. We conduct additional evaluation on the various stages of our pipeline. For YOLOv5, we held out 20 frames for evaluation and found that the mean average precision for predicting both the ocular globe and the optic nerve was 99.5\%. For the U-Net model the mean average error for the predicted width is 1.04 mm.
A visualization of a predicted R2U-Net mask can be seen in Figure \ref{fig:stacked_nerve}.

% 8.829 pixels / 8.52

% \subsection{Qualitative Results}
In addition to these automated evaluation metrics, we also had one of our SMEs qualitatively assess a random assortment of 20 test images from our R2U-Net model, similar to Figure \ref{fig:unet_ex}. We discarded 5 examples that the SME was unable to interpret, leaving 15 images in total. For each example, the SME was tasked with providing a binary response to 2 questions: ``Does the predicted angle correctly correspond to the angle of the nerve?" and ``Does the predicted width correctly correspond to the width of the nerve at the measurement location?". For the first question, the SME answered `yes' for 13/15 of the images, indicating that the model is able to correctly predict the angle in 87\% of the generated cases. For the second question, the SME answered `yes' for 12/15 of the images, indicating that the model correctly predicts the width in 80\% of the generated images.

\subsection{Discussion}
As discussed in Section \ref{sec:intro}, there are a variety of target criteria that must be met in order for an AI system to be suitable for deployment in our application setting. The first of these criteria focused on developing models that are interpretable and easy to use so that they can support the medical examiner throughout the procedure. Our expert-driven decomposition approach to model design allows us to surface relevant information at intermediate states of the model, which would otherwise not be possible in an end-to-end system. In our sparse coding model, some additional feedback is provided to the examiner in the form of a bounding box around the nerve. Our R2U-Net system surfaces even more information (see Figure \ref{fig:unet_ex}). A blue line extends from the posterior aspect of the ocular globe to the 3mm mark in the nerve, while also indicating the angle in which the optic nerve approaches the retinal plane, the red dot indicates the point of measurement, and the orange line corresponds to the predicted width. These artifacts accurately represent the actual values that the model is using to make its prediction, rather than some approximation generated by a post hoc method. The information provided by our system can serve many different purposes depending on the target user. If the user is an expert, it may simply speed up image interpretation. If the user is a non-expert, such as a battlefield medic, it may fill gaps in their understanding and enable them to successfully collect a measurement.

\section{Conclusion}
In this paper we presented two systems for detecting elevated ICP in an ultrasound video by measuring the ONSD. We created our systems to mirror the procedure that SMEs use to complete the task. This allows us to surface relevant information to the examiner throughout the procedure, supporting them rather than supplanting them. Additionally, we selected model components that are able to be executed on mobile platforms, creating a clear path for potential future deployment. We conducted a number of experiments that demonstrate our systems are able to achieve strong results, outperforming larger, pre-trained baseline architectures, and further validated the utility of our top performing model by having a SME manually evaluate the quality of its predictions.

\section{Conflicts of Interest}
This work was supported by Defense Advanced Research Projects Agency (DARPA), SPARTACUS-X: Sparse Coding and Extraction of Ultrasound Knowledge for Explainable POCUS AI, HR00112190076, \$999,999, May 2021 - October 2022. (PI Christopher MacLellan with Co-PIs Rosina Weber and Edward Kim)

\section{Compliance with Ethical Standards}
The study protocol was reviewed by Drexel University's Institutional Review board as well as the Human Research Protection Office at the U.S. Army Medical Research and Development Command. Both bodies determined that our study "is not human subjects research as defined by DHHS or FDA regulations".

\bibliographystyle{IEEEbib}
\bibliography{main}

\end{document}